\documentclass[preprint,sort&compress]{elsarticle}
\usepackage{graphicx}
\usepackage{amsmath}
\usepackage{amsfonts}
\usepackage{amssymb}
\usepackage{epsfig}
\usepackage{hyperref}

\def \be {\mathbf{E}}

\def \br {\mathbf{r}}

\def \bx {\mathbf{x}}
\def \by {\mathbf{y}}
\def \bq {\mathbf{q}}

\def \cf {C_F}
\def \nc {N_c}

\def \bp {\mathbf{p}}

\newcommand{\Tint}[1]{{\hbox{$\sum$}\!\!\!\!\!\!\!\int\,}_{\!\!\!\!\raise-0.9ex\hbox{$\scriptstyle{#1}$}}}

\def \mbq {\vert\bq\vert}

\def\siml{{\ \lower-1.2pt\vbox{\hbox{\rlap{$<$}\lower6pt\vbox{\hbox{$\sim$}}}}\ }}
\def\simg{{\ \lower-1.2pt\vbox{\hbox{\rlap{$>$}\lower6pt\vbox{\hbox{$\sim$}}}}\ }}

\def \be {\mathbf{E}}

\def \br {\mathbf{r}}

\def \bx {\mathbf{x}}
\def \by {\mathbf{y}}
\def \bq {\mathbf{q}}

\def \bfnabla {\boldsymbol{\nabla}}

\def \bp {\mathbf{p}}

\def \mbq {\vert\bq\vert}
\def \mbp {\vert\bp\vert}
\def \mbr {\vert\br\vert}

\def \als {\alpha_{\mathrm{s}}}

\def \m2   {\mu^{2 \epsilon}}

\newcommand{\order}[1]{\mathcal{O}\left(#1\right)}

\def\siml{{\ \lower-1.2pt\vbox{\hbox{\rlap{$<$}\lower6pt\vbox{\hbox{$\sim$}}}}\ }}
\def\simg{{\ \lower-1.2pt\vbox{\hbox{\rlap{$>$}\lower6pt\vbox{\hbox{$\sim$}}}}\ }}
\def\lqcd{\Lambda_\mathrm{QCD}}

\def\nn {\nonumber}

\begin{document}
\title{Thermal width and gluo-dissociation of quarkonium in pNRQCD}
\author[tum]{Nora Brambilla}
\author[tum]{Miguel \'Angel Escobedo}
\author[tum,exc]{Jacopo Ghiglieri}
\author[tum]{Antonio Vairo}
\address[tum]{Physik-Department, Technische Universit\"at M\"unchen,
James-Franck-Str. 1, 85748 Garching, Germany}
\address[exc]{Excellence Cluster Universe, Technische Universit\"at M\"unchen, 
Boltzmannstr. 2, 85748, Garching, Germany}

\begin{abstract}
The thermal width of heavy-quarkonium bound states in a quark-gluon
plasma has been recently derived in an effective field theory approach.
Two phenomena contribute to the width: the Landau damping phenomenon and 
the break-up of a colour-singlet bound state into a colour-octet
heavy quark-antiquark pair by absorption of a thermal gluon.
In the paper, we investigate the relation between the singlet-to-octet 
thermal break-up and the so-called gluo-dissociation, a mechanism for quarkonium 
dissociation widely used in phenomenological approaches.
The gluo-dissociation thermal width is obtained by convoluting 
the gluon thermal distribution with the cross section of a gluon and 
a $1S$ quarkonium state to a colour octet quark-antiquark state in vacuum, 
a cross section that at leading order, but neglecting colour-octet effects,  
was computed long ago by Bhanot and Peskin.
We will, first, show that the effective field theory framework provides 
a natural derivation of the gluo-dissociation factorization formula at leading order, 
which is, indeed, the singlet-to-octet thermal break-up expression.
Second, the singlet-to-octet thermal break-up expression will allow us to improve the Bhanot--Peskin cross section 
by including the contribution of the octet potential, which amounts to include 
final-state interactions between the heavy quark and antiquark. 
Finally, we will quantify the effects due to final-state interactions
on the gluo-dissociation cross section and on the quarkonium thermal width.
\end{abstract}

\begin{keyword}
Quarkonium, finite temperature, thermal width, gluo-dissociation, singlet-to-octet break-up
\end{keyword}
\maketitle

\section{Introduction}
Quarkonium suppression has been suggested long ago as a hard probe of 
the medium produced in heavy-ion collisions \cite{Matsui:1986dk}.
This hypothesis has been widely investigated,
both theoretically and experimentally, in the past 25 years
\cite{Brambilla:2004wf,Brambilla:2010cs}. 
The early theoretical arguments were based on the expectation that above the
deconfinement temperature the linear, confining part of the heavy
quark-antiquark ($Q\overline{Q}$) potential would vanish and the
Coulomb part at short distances would be replaced by a screened Yukawa
(or Debye) potential that can support only a limited number of bound
states. Since the screening (Debye) mass depends on the temperature, 
heavy-quarkonium states were thought to provide a thermometer of the medium.

In the last few years, significant progress has been made in deriving
the heavy $Q\overline{Q}$ potential from QCD in a rigorous and systematic way. 
The real-time static potential was first calculated for large
temperatures, $T \gg 1/r \simg m_D$, where $r$ is the quark-antiquark distance and 
$m_D$ is the Debye mass, in \cite{Laine:2006ns,Laine:2007gj,Burnier:2007qm,Beraudo:2007ky}. 
For a wider range of temperatures, an effective field theory (EFT) study of
non-relativistic bound states at finite temperature has been carried
out for QED in \cite{Escobedo:2008sy,Escobedo:2010tu,Escobedo:2011ie}
and for QCD in \cite{Brambilla:2008cx,Brambilla:2010vq,Brambilla:2011mk}.  
Most importantly, in the same framework, also the quarkonium thermal width 
has been calculated. Two mechanisms, at least, have been identified 
as responsible for it: the \emph{Landau damping} phenomenon \cite{Laine:2006ns} and the
\emph{singlet-to-octet thermal break-up} \cite{Brambilla:2008cx}. 
In the former, the virtual gluons that are exchanged between the $Q\overline{Q}$ 
pair scatter off the light constituents of the medium, 
whereas in the latter the colour-singlet bound state absorbs 
a gluon from the medium and turns into a colour-octet state.

There exists, however, a large literature where the quarkonium behaviour 
in a medium is studied on a phenomenological basis
(see e.g. \cite{Mocsy:2007jz,Rapp:2008tf,Kluberg:2009wc}).
The thermal decay width is obtained by convoluting scattering cross sections computed at
$T=0$ with thermal distributions for the incoming light partons. 
At least two scattering processes have been considered:
\emph{gluo-dissociation} and \emph{quasi-free dissociation}.
It is then natural to ask, if and to what extent
gluo-dissociation and quasi-free dissociation agree with the 
singlet-to-octet break-up and the Landau damping widths derived from EFTs. 

In this letter, we will deal with gluo-dissociation 
\cite{Kharzeev:1994pz,Xu:1995eb} (see
\cite{Xu:2007yu,Polleri:2003kn,Patra:2004wf,Patra:2005yg,Patra:2005bi,Wong:2004zr,Arleo:2004ge,Thews:2005vj,Grandchamp:2005yw,Park:2007zza,Liu:2009wza,Qu:2009sk,Zhou:2009vz,Song:2010ix,Zhao:2010nk,Liu:2010ej,Uphoff:2011fu,Mandal:2011jx}
for some recent literature and \cite{Rapp:2008tf,Rapp:2009my,Kluberg:2009wc} for reviews). 
The mechanism of quasi-free dissociation \cite{Grandchamp:2001pf}
and its relation with the EFT framework will be dealt with elsewhere \cite{preparation}.
The process underlying gluo-dissociation is the same that gives rise to the
singlet-to-octet thermal break-up width in the EFT framework. 
The cross section for gluon absorption by a colour-singlet $1S$ state was
computed in 1979 by Bhanot and Peskin (BP) in Refs.~\cite{Peskin:1979va,Bhanot:1979vb}, 
where the contribution from the final-state interactions was neglected by considering 
the large number of color, $\nc$, limit.
In the present work, we will 
\emph{(i)} prove at leading order the factorization formula 
for quarkonium gluo-dissociation and establish under which conditions it holds, 
hence, show that the gluo-dissociation thermal width coincides with the singlet-to-octet 
thermal break-up width, 
\emph{(ii)} improve the gluo-dissociation 
cross section by including the contribution of the octet potential, which amounts to include 
$Q\overline{Q}$ final-state interactions, and, finally, 
\emph{(iii)} assess quantitatively the effects due to final-state interactions
on the gluo-dissociation cross section and on the thermal width.

The letter is organized in the following way.
In the next section, we will recall some basics on gluo-dissociation. 
In Sec.~\ref{gluo_EFT}, we will derive the gluo-dissociation factorization formula in an EFT framework, 
indeed showing that it coincides with the singlet-to-octet thermal break-up
expression, first derived in \cite{Brambilla:2010vq}. 
We will also show that the gluo-dissociation cross section agrees with the BP cross section in the 
large $\nc$ limit. In Sec.~\ref{sec_octet}, we will improve the BP cross section 
by including final-state interactions between the $Q\overline{Q}$ pair in a colour-octet state. 
Finally, in Sec.~\ref{sec_concl}, we will draw some conclusions. 
The results presented here are also part of the Ph.D. thesis \cite{PHDJacopo}.

\section{Gluo-dissociation}
\label{secgd}
Gluo-dissociation stands for the process $g+\Phi(1S)\to (Q\overline{Q})_8$,
where a quarkonium $1S$ state, $\Phi(1S)$, absorbs a gluon and becomes 
an unbound $Q\overline{Q}$ pair in a colour-octet state, $(Q\overline{Q})_8$.
In the literature, it has been assumed that convolving 
the in vacuum gluo-dissociation cross section, $\sigma_{1S}$, 
with the thermal distribution of the gluons provides the 
gluo-dissociation thermal width, $\Gamma_{1S}$.\footnote{
There exist papers, such as \cite{Wong:2004zr,Arleo:2004ge}, where 
finite-temperature effects in the cross section are included 
by considering colour-singlet wavefunctions derived from potential models. 
On the thermal distribution side, hydrodynamical and anisotropic effects 
have been considered in \cite{Patra:2004wf,Patra:2005yg,Patra:2005bi,Mandal:2011jx}.}  
Specifically, one writes (see for instance Eq. (23) of \cite{Rapp:2008tf})
\begin{equation}
\Gamma_{1S}=\int_{\mbq \ge \vert E_{1S}\vert}\frac{d^3q}{(2\pi)^3}n_\mathrm{B}(\mbq)\,\sigma_{1S}(\mbq)\,,
\label{gluodiss}
\end{equation}
where $n_\mathrm{B}(x)\equiv(e^{x/T}-1)^{-1}$ is the Bose--Einstein distribution, 
$E_{1S}$ the binding energy of the quarkonium $1S$ state, 
and we have assumed the bound state and the bath to be at rest. 

In \cite{Peskin:1979va,Bhanot:1979vb}, the gluo-dissociation cross section was 
calculated at leading order, under the following assumptions:
\emph{(1)} the quarkonium $1S$ state, $\Phi(1S)$, is Coulombic;
\emph{(2)} in an operator product expansion framework, 
the gluon-quarkonium interaction is taken at leading order, 
which corresponds to a chromoelectric dipole interaction; 
\emph{(3)} the (repulsive) octet potential is neglected, which is
tantamount to neglecting final-state interactions.
Theoretically the last assumption may be realized by taking the large-$\nc$ limit.
In this limit, the colour-singlet Coulomb potential, i.e. 
the potential between a  $Q\overline{Q}$ pair in a colour-singlet configuration, which is (at leading order)
$V^{(0)}_s=-\cf\als/r$ with $\cf=(\nc^2-1)/(2\nc)$, becomes  $V^{(0)}_{s\, {\rm BP}}=-\nc\als/(2r)$, 
whereas the colour-octet Coulomb potential, i.e. the potential between a  $Q\overline{Q}$ pair in a colour-octet configuration, 
which is $V^{(0)}_o= \als/(2\nc r)$, vanishes.
The large-$\nc$ limit also modifies the Bohr radius from $a_0=2/(m\cf\als)$ to $a_\mathrm{BP}=4/(m\nc\als)=4/(3m\als)$ and the
absolute value of the binding energy (at leading order) from $\vert E_1\vert=m\cf^2\als^2/4$ 
to $\epsilon_{1,\mathrm{BP}}=1/(ma_\mathrm{BP}^2)=9\,m\als^2/16$. 
The Bhanot--Peskin gluo-dissociation cross section, $\sigma_{1S,\mathrm{BP}}(\mbq)$,
as a function of the gluon momentum $\bq$, then reads 
\begin{eqnarray}
\sigma_{1S,\mathrm{BP}}(\mbq) 
&=&\frac{2^9\pi\als}{9}\frac{\epsilon_{1,\mathrm{BP}}^{5/2}}{m}\frac{(\mbq-\epsilon_{1,\mathrm{BP}})^{3/2}}{\mbq^5}.
\label{peskinrewrite}
\end{eqnarray}
We note that, in the above formula, an overall colour factor $\cf = 4/3$ has been kept unexpanded when performing 
the large $\nc$ limit. We call $\Gamma_{1S,\mathrm{BP}}$ the corresponding gluo-dissociation thermal width.

The cross section \eqref{peskinrewrite} is averaged over the 2 polarizations and the 8 colours of the initial gluon. 
Hence, when inserted in the decay width formula, Eq. \eqref{gluodiss}, the Bhanot--Peskin cross section should be multiplied 
by a factor 16: $\sigma_{1S}(\mbq) \approx 16 \times \sigma_{1S,\mathrm{BP}}(\mbq)$.
The factor is explicitly included in Eq. (4) of \cite{Park:2007zza} and also the
authors of \cite{Zhao:2010nk} multiply the BP cross section by 16 when using 
Eq.~\eqref{gluodiss}.\footnote{Private communications from Xingbo Zhao
are acknowledged.}

\section{Effective field theory}
\label{gluo_EFT}
The EFT approach is based on the hierarchies of non-relativistic 
and thermal scales typical of quarkonium in a quark-gluon plasma.
The hierarchy of non-relativistic scales follows from the fact that the heavy quark 
has a velocity in the centre-of-mass frame that is  $v \ll 1$; 
the hierarchy is then $m \gg mv \gg mv^2$, where $m$ 
is the heavy-quark mass, $mv$ is the scale of the typical momentum transfer in 
or inverse radius of the bound state and $mv^2$ is the scale of the 
typical energy. For a Coulombic bound state, such as the bottomonium ground state
likely is, it holds that $mv \sim m\als \gg \lqcd$ and also that $E\sim m\als^2\simg\lqcd$. 
The hierarchy of thermal scales is $T \gg m_D$, where $T$ is the temperature 
of the quark-gluon plasma and $m_D$ the Debye mass.\footnote{
The right temperature scale is rather $\pi T$, or multiples thereof, than $T$.
This is, however, relevant, only when quantifying the different 
energy scales in the system; hence, we will drop the factor $\pi$ from qualitative considerations.
Effects due to the magnetic mass are suppressed and do not contribute to the considered accuracy.}

The relative size of non-relativistic and thermal scales depends on the medium and on the quarkonium state.
In the following, we will adopt the hierarchy  
\begin{equation}
m \gg mv \sim m\als \gg T \sim mv^2\sim m\als^2 \gg m_D,\lqcd\,,
\label{scales}
\end{equation}
which is meant to include also the regions $mv \sim m\als \gg T \simg mv^2\sim m\als^2$ and 
$mv^2\sim m\als^2 \simg T \gg m_D,\lqcd$, although this last one is of less phenomenological
impact for the thermal width is exponentially suppressed \cite{Brambilla:2008cx}.
The hierarchy \eqref{scales} is more general than the one 
analyzed in \cite{Brambilla:2010vq}, where we required $mv \sim m\als \gg T \gg mv^2 \sim m\als^2$.
It was argued in \cite{Vairo:2010bm} that this hierarchy may be the relevant one 
for $\Upsilon(1S)$ produced in heavy-ion collisions at the LHC.\footnote{
We refer to \cite{Chatrchyan:2011pe} for the most recent 
CMS measurements on the suppression of the $\Upsilon$ family. 
A phenomenological analysis of the data that includes the effects of the Landau-damping width
can be found in \cite{Strickland:2011mw}.}
In \cite{Brambilla:2010vq}, the spectrum and width of quarkonia were computed up to order $m\als^5$.

EFTs suitable to describe quarkonium in a medium are constructed by subsequently integrating out 
high-energy scales in \eqref{scales}. 
Integrating out modes that scale like $m$ and $m\als$ leads respectively to non-relativistic QCD
(NRQCD) \cite{Caswell:1985ui,Bodwin:1994jh} and potential
non-relativistic QCD (pNRQCD) \cite{Pineda:1997bj,Brambilla:1999xf}. Since the temperature is much
smaller than both $m$ and $m\als$, it can be set to zero in the
matching and both Lagrangians are the same as at zero temperature.
The pNRQCD Lagrangian, in particular, reads
\begin{eqnarray}
&& \hspace{-1.5cm}
{\cal L}_{\textrm{pNRQCD}} = - \frac{1}{4} F^a_{\mu \nu} F^{a\,\mu \nu} + \sum_{i=1}^{n_f}\bar{q}_i\,iD\!\!\!\!/\,q_i
\nonumber\\ 
&&\hspace{0.5cm}
+ \int d^3r \; {\rm Tr} \,  
\Bigl\{ {\rm S}^\dagger \left[ i\partial_0 - h_s \right] {\rm S} 
+ {\rm O}^\dagger \left[ iD_0 -h_o \right] {\rm O} 
\nonumber \\
&& \hspace{1.8cm}
+ V_A\, \left( {\rm O}^\dagger \br \cdot g\be \,{\rm S} + \textrm{H.c.} \right)
+ \frac{V_B}{2} {\rm O}^\dagger \left\{ \br\cdot g\be \,, {\rm O}\right\} 
+ \dots\,\Bigr\} .
\label{pNRQCD}	
\end{eqnarray}
The fields $\mathrm{S}=S\,\mathbf{1}_c/\sqrt{N_c}$ and $\mathrm{O}=O^a\,T^a/\sqrt{T_F}$ are the $Q\overline{Q}$ 
colour-singlet and colour-octet fields respectively, $n_f$ is the number of light quarks, $q_i$, $T_F=1/2$,
$\be$ is the chromoelectric field,  $iD_0 \mathrm{O} =i\partial_0 \mathrm{O} - gA_0 \mathrm{O} + \mathrm{O} gA_0$ 
and H.c. stands for Hermitian conjugate. The trace is over colour and spin indices. 
Gluon fields depend only on the centre-of-mass coordinate and on time;
this is a consequence of having multipole expanded the gluon fields in the quark-antiquark relative
distance $r$. The dots in the last line stand for higher-order terms in $r$ and $1/m$.
We note that, as in the BP approach, the leading gluon-quarkonium interaction is a 
chromoelectric dipole interaction.

The dependence on the scales $m$ and $m\als$ is encoded in the Wilson coefficients; $V_A$
and $V_B$ are at leading order $V_A=V_B=1$, whereas the singlet and octet Hamiltonians have the form 
($\bp\equiv-i\nabla_\br$) 
\begin{equation}
h_{s,o}=\frac{\bp^2}{m} +V^{(0)}_{s,o} +\frac{V^{(1)}_{s,o}}{m}+\frac{V^{(2)}_{s,o}}{m^2}+\ldots\,.
\label{sinoctham}
\end{equation}
The dots stand for higher-order terms in the $1/m$ expansion.
The first two terms in the right-hand side, which are the kinetic energy 
and the static potential, constitute the leading-order Hamiltonian.
The singlet and octet leading-order Hamiltonians explicitly read 
\begin{equation}
h^{(0)}_s =\frac{\bp^2}{m}-\cf\frac{\als}{r},\qquad 
h^{(0)}_o =\frac{\bp^2}{m}+\frac{1}{2\nc}\frac{\als}{r}.
\label{leadinghams}
\end{equation}
The spectrum of $h_s^{(0)}$ is made by the (QCD) Bohr levels $E_n = -m\cf^2\als^2/(4n^2)$, 
whereas the octet potential is repulsive and does not support bound states but 
a continuum of scattering states. Note that, in the non-relativistic EFT power counting, 
both the kinetic energy and the static potential scale like $m\als^2$. 
Therefore, neglecting the octet potential, as done in the BP calculation, 
is a sensible approximation only in the large $\nc$ limit.

\begin{figure}[ht]
\begin{center}
\includegraphics[width=7.5cm]{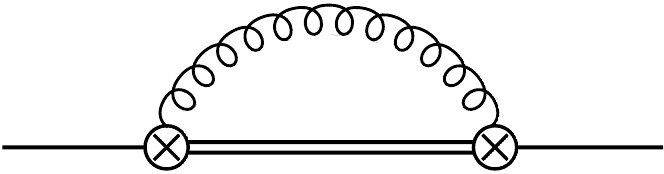}
\end{center}
\caption{The leading heavy-quarkonium self-energy diagram in pNRQCD. 
The single line is a singlet $Q\overline{Q}$ propagator, the double line an octet $Q\overline{Q}$ propagator, 
the curly line a gluon and the vertices are chromoelectric dipole vertices. 
The imaginary part, obtained by cutting this diagram, gives the singlet-to-octet break-up thermal width.}
\label{fig:leading}
\end{figure}

We now set out to compute the singlet-to-octet break-up amplitude and width within pNRQCD. 
We refer to \cite{Brambilla:2008cx,Brambilla:2010vq} for details regarding pNRQCD at finite temperature 
in the real-time formalism.\footnote{Since we do not require here $T \gg m\als^2$, a difference with \cite{Brambilla:2010vq} 
is that we avoid integrating out $T$ and constructing the intermediate EFT called pNRQCD${}_\mathrm{HTL}$ \cite{Vairo:2009ih}.}
The singlet-to-octet break-up thermal width is given by the imaginary part of Fig.~\ref{fig:leading}. 
The amplitude $\Sigma(E)$ of that diagram reads
\begin{eqnarray}
\Sigma(E) \!\!&=&\!\! - i g^2 \, C_F \, \frac{2}{3}r^i \int \frac{d^4k}{(2\pi)^4} \frac{i}{E-h^{(0)}_o-k_0 +i\eta}
\nonumber\\
&&\hspace{2.5cm}\times\left[k_0^2\, 2\pi n_{\rm B}(\vert k_0\vert)\, \delta(k_0^2-k^2)\right]r^i\,,
\label{widthdef}
\end{eqnarray}
where $E$ is the energy of the incoming  $Q\overline{Q}$  singlet and the expression in square brackets is the thermal part 
of the chromoelectric correlator. Since $T\sim E \gg m_D$, bare propagators can be used. 
The number of dimensions has been set to 4, the integral being convergent. Evaluating it yields for the imaginary part
\begin{equation}
\label{ampliwidth}
 \mathrm{Im}\, \Sigma(E) = -\frac{g^2\cf}{6\pi}r^i\left\vert E-h_o^{(0)}\right\vert^3n_\mathrm{B}(\vert E-h_o^{(0)}\vert)r^i\,.
\end{equation}
The singlet-to-octet break-up width for the $1S$ state then reads
\begin{equation}
\label{width}
\Gamma_{1S} = -2\langle 1S\vert\mathrm{Im}\,\Sigma(E_{1S})\vert 1S\rangle
=\frac{g^2\cf}{3\pi}\langle1S\vert r^i\left\vert E_1-h_o^{(0)}\right\vert^3n_\mathrm{B}(\vert E_1-h_o^{(0)}\vert)r^i\vert 1S\rangle\,,
\end{equation}
where $\langle {\bf r} \vert 1S\rangle=1/(\sqrt{\pi}a_0^{-3/2})\exp(-r/a_0)$ is the
Coulomb $1S$ wavefunction. The corresponding expression in the static limit 
was obtained in \cite{Brambilla:2008cx} and in QED, for the hydrogen atom, in \cite{Escobedo:2008sy}. 

The difficulty in the evaluation of Eq. \eqref{width} lies in the Bose--Einstein distribution and in its
nontrivial dependence on $h_o^{(0)}$. 
In Ref.~\cite{Brambilla:2010vq}, having instead assumed $T\gg m\als^2$, we could expand the Bose--Einstein
distribution as $n_\mathrm{B}(\vert E_1-h_o^{(0)}\vert) = T/\vert E_1-h_o^{(0)} \vert-1/2+\ldots$, obtaining 
up to corrections of order $m\als^5\, E_1/T$
\begin{eqnarray}
\nonumber
\Gamma_{1S} \!\!  &=& \!\! 
\frac{T\als^3}{3}\left( 4C_F^3 + 4C_F^2\nc + C_F\nc^2 \right)
\\
&& 
- \frac{mC_F^2\als^5}{24}\left( 16C_F^3 + 20C_F^2\nc + 8C_F\nc^2 + \nc^3\right).
\label{analyticwidth}
\end{eqnarray}
The terms in the first line are the leading ones and are linear in the temperature.\footnote{
The linear behaviour of the thermal width has been recently investigated and found consistent with 
lattice data in \cite{Aarts:2011sm}.}

Without expanding the Bose--Einstein distribution, the matrix element in Eq.~\eqref{width} can be
evaluated analogously to how the QCD Bethe logarithms, which contribute
to the quarkonium spectrum at order $m\als^5$, have been dealt with in
\cite{Kniehl:1999ud,Kniehl:2002br}, i.e. by inserting a complete set of octet states.
Octet states are labeled by their energy and angular momentum quantum numbers 
and obey $h_o^{(0)}\vert pll_z \rangle=(p^2/m) \vert pll_z\rangle$. It is convenient to introduce 
an arbitrary unit vector $\hat{\bp}$ and define a state 
$\displaystyle \vert \bp l\rangle \equiv (4\pi/p) \sum_{l_z} \vert pll_z\rangle \langle ll_z\vert \hat{\bp}\rangle$, 
where $\langle ll_z\vert \hat{\bp}\rangle =  Y_l^{l_z}(\hat{\bp})^*$ is a spherical harmonics.
A suitable normalization of the states  $\vert \bp l\rangle$ is 
\begin{equation}
\label{defnormp}
\sum_l \int\frac{d^3p}{(2\pi)^3}\langle\bx\vert\bp l\rangle\langle\bp l\vert\by\rangle=\delta^3(\bx-\by)\,.
\end{equation}
Inserting \eqref{defnormp} into \eqref{width} gives
\begin{eqnarray}
\Gamma_{1S} \!\!&=&\!\! \int_{\mbq \ge |E_1|}\frac{d^3q}{(2\pi)^3}\,n_\mathrm{B}\left(\mbq\right)
\nn\\
&& \hspace{0.8cm}
\times \frac{g^2\cf}{3\pi} \frac{m^{3/2}\mbq \sqrt{\mbq+E_1}}{2}
\left\vert\langle1S\vert {\bf r}\vert \bp 1\rangle\right\vert^2 \Big\vert_{\mbp=\sqrt{m(\mbq+E_1)}}\,,
\label{widthinsert}
\end{eqnarray}
where we have used that $\left|E_1-h_o^{(0)}\right|^3n_\mathrm{B}(|E_1-h_o^{(0)}|)$ is analytic in $h_o^{(0)}$ 
and made explicit that $\langle1S\vert {\bf r}$ projects on a $l=1$ state.
Equation \eqref{widthinsert} provides, at leading order in the EFT power counting, 
the decay width associated to the quarkonium singlet-to-octet thermal break-up, which is 
the dominant decay process in the situation \eqref{scales}.
Quarkonium singlet-to-octet break up describes, at the order we are working, the same 
process of quarkonium gluo-dissociation, and the thermal decay width \eqref{widthinsert}
may be identified with the gluo-dissociation thermal width.
In the following, we will show that Eq. \eqref{widthinsert} satisfies, indeed, the properties of the 
gluo-dissociation width presented in Sec.~\ref{secgd}.

First, since the thermal decay width  \eqref{widthinsert} is expressed 
as a convolution of the gluon Bose--Einstein distribution and a function of the gluon 
momentum, this proves, at leading order, the factorization formula \eqref{gluodiss}. 
It also allows the identification 
\begin{equation}
\sigma_{1S}(\mbq) = 
\frac{g^2\cf}{3\pi} \frac{m^{3/2}\mbq \sqrt{\mbq+E_1}}{2}
\left\vert\langle1S\vert {\bf r}\vert \bp 1\rangle\right\vert^2 \Big\vert_{\mbp=\sqrt{m(\mbq+E_1)}}\,,
\label{sigma}
\end{equation}
for $\mbq \ge |E_1|$.
We note that, in an alternative derivation, we could have made use of 
cutting rules at finite temperature \cite{Kobes:1986za} for the imaginary part 
of the diagram in Fig.~\ref{fig:leading}. This would have led again  
to the factorization formula \eqref{gluodiss} and to the identification of 
$\sigma_{1S}$ with a $T=0$ cross section. However, the factorization formula 
\eqref{gluodiss} is not expected to hold at higher orders.

Finally, in order to reproduce the BP gluo-dissociation cross section \eqref{peskinrewrite} 
from \eqref{sigma}, we evaluate the dipole matrix element squared
$\left\vert\langle1S\vert \br\vert \bp 1 \rangle\right\vert^2$ in the
absence of the octet potential. This is tantamount to using 
plane waves for the octet wave functions: 
$\displaystyle \sum_l \langle\bx\vert\bp l\rangle\langle\bp l\vert\by\rangle = e^{i\bp\cdot(\bx-\by)}$.
The matrix element squared then becomes the
square of the derivative of the momentum-space wavefunction and reads
\begin{equation}
\label{peskinmatrix}
\left\vert\langle1S\vert \br\vert \bp 1\rangle\right\vert^2\,{\buildrel\nc\to\infty\over\longrightarrow}\,
\left\vert \bfnabla_{\bp}\langle \bp\vert 1S\rangle\right\vert^2
=\frac{2^{10}\pi a_\mathrm{BP}^7\mbp^2}{(1+a_\mathrm{BP}^2\mbp^2)^6}.
\end{equation}
Plugging this into Eq.~\eqref{sigma} and replacing $E_1\to-\epsilon_{1,\mathrm{BP}}$, $C_F=4/3$, we obtain
\begin{eqnarray}
\label{eftcrosslargenc}
\sigma_{1S} (\mbq) \!\!&{\buildrel\nc\to\infty\over\longrightarrow}& \!\!
16\frac{2^9\pi\als}{9}\frac{\epsilon_{1,\mathrm{BP}}^{5/2}}{m}\frac{(\mbq-\epsilon_{1,\mathrm{BP}})^{3/2}}{\mbq^5}
= 16 \,\sigma_{1S,\mathrm{BP}}(\mbq)\,,
\\
\label{peskinEFT}
\Gamma_{1S}\!\!&{\buildrel\nc\to\infty\over\longrightarrow}&\!\! 
\int_{\mbq \ge \epsilon_{1,\mathrm{BP}}}
\frac{d^3q}{(2\pi)^3} n_\mathrm{B}(\mbq)\, 16 \,\sigma_{1S,\mathrm{BP}}(\mbq)
= \Gamma_{1S,\mathrm{BP}}.
\end{eqnarray} 
Equation \eqref{eftcrosslargenc} reproduces the BP cross section \eqref{peskinrewrite}; 
as already discussed, the factor 16 accounts for the 2 polarizations and 8 colours of the gluon.
Therefore, the EFT computation in the large-$\nc$ limit leads naturally to the 
BP factorization and cross section formulas for quarkonium gluo-dissociation in a medium.
We now set out to include the octet potential in the calculation, thereby also quantifying the
approximation introduced by neglecting it.

\section{Colour-octet effects}
\label{sec_octet}
The calculation of the dipole matrix element squared
$\left\vert\langle1S\vert \br \vert \bp 1\rangle\right\vert^2$, 
when octet potential contributions are included, 
is more involved, and requires the explicit integration over the
continuum octet wavefunctions. Coulombic wavefunctions in the continuum
region $\vert pll_z\rangle$ can be found in \cite{abramovitz+stegun}; 
$l=1$ octet wavefunctions $ \vert\bp 1\rangle$ can be found in \cite{Kniehl:1999ud,Kniehl:2002br}.
After correcting some typos, they read
\begin{eqnarray}
\langle \br \vert\bp 1\rangle \!\!&=&\!\!  e^{i(\pi/2-\delta_1)}
\sqrt{2\pi} \bp\cdot\br 
\sqrt{\frac{\rho\left(1+\frac{\rho^2}{a_0^2 \mbp^2}\right)}{a_0\mbp\left(e^{\frac{2 \pi  \rho }{a_0 \mbp}}-1\right)}}
e^{i\mbp\mbr}\, 
\nn\\
&& \hspace{3cm} \times \hbox{$ $}_1F_1(2+i\rho/(a_0\mbp);4;-i2\mbp\mbr)\,,
\end{eqnarray}
where $\hbox{$ $}_1F_1$ is the confluent hypergeometric function, $\delta_1$ is the $l=1$ Coulomb phase  
and $\rho\equiv 1/(\nc^2-1)$. The matrix element squared is then 
\begin{equation}
\label{penin}
\left\vert\langle1S\vert \br\vert \bp 1\rangle\right\vert^2=
\frac{512 \pi ^2 \rho  (\rho +2)^2a_0^6\mbp \left(1+\frac{\rho ^2}{a_0^2 \mbp^2}\right)
e^{\frac{4 \rho}{a_0 \mbp}  \arctan(a_0 \mbp)}}{\left(e^{\frac{2 \pi  \rho }{a_0 \mbp}}-1\right) \left(1+a_0^2\mbp^2\right)^6}.
\end{equation}
It is easily seen that the $\nc\to\infty$ ($\rho\to0$) limit of this equation gives back Eq.~\eqref{peskinmatrix}. 
Plugging the matrix element into Eq.~\eqref{sigma} yields
\begin{equation}
\sigma_{1S}(\mbq)=\frac{\als\cf}{3} 2^{10} \pi^2  \rho  (\rho +2)^2 \frac{E_1^{4}}{m\mbq^5}
\left(t(\mbq)^2+\rho ^2	\right)\frac{\exp\left(\frac{4 \rho}{t(\mbq)}  \arctan
\left(t(\mbq)\right)\right)}{ e^{\frac{2 \pi  \rho}{t(\mbq)} }-1}\,,
\label{crossEFT}
\end{equation}
where $t(\mbq)\equiv\sqrt{\mbq/\vert E_1\vert-1}$. 
The limit  $\nc\to\infty$ ($\rho\to0$) gives back Eq.~\eqref{eftcrosslargenc}.\footnote{
\label{foot}
The $2S$ cross section reads 
\begin{eqnarray}
\sigma_{2S}(\mbq)\!\!&=&\!\!
\frac{\als\cf}{3} 2^{13} \pi ^2 \rho \frac{E_2^4}{ m \mbq^7}
\left[2 E_2 \left(2 \rho^2+5 \rho +3\right)+\mbq (\rho +2)\right]^2
\left( t_2(\mbq)^2 + 4 \rho^2 \right) 
\nn\\
&&\times 
\exp\left(\frac{8 \rho}{t_2(\mbq)} \arctan \left(t_2(\mbq)\right)\right)
\left[e^{\frac{4 \pi \rho}{t_2(\mbq)}}-1\right]^{-1}\,,
\nn
\end{eqnarray}
where $t_2(\mbq)\equiv\sqrt{\mbq/\vert E_2\vert-1}$.
} 

\begin{figure}[ht]
\begin{center}
\includegraphics[width=12cm]{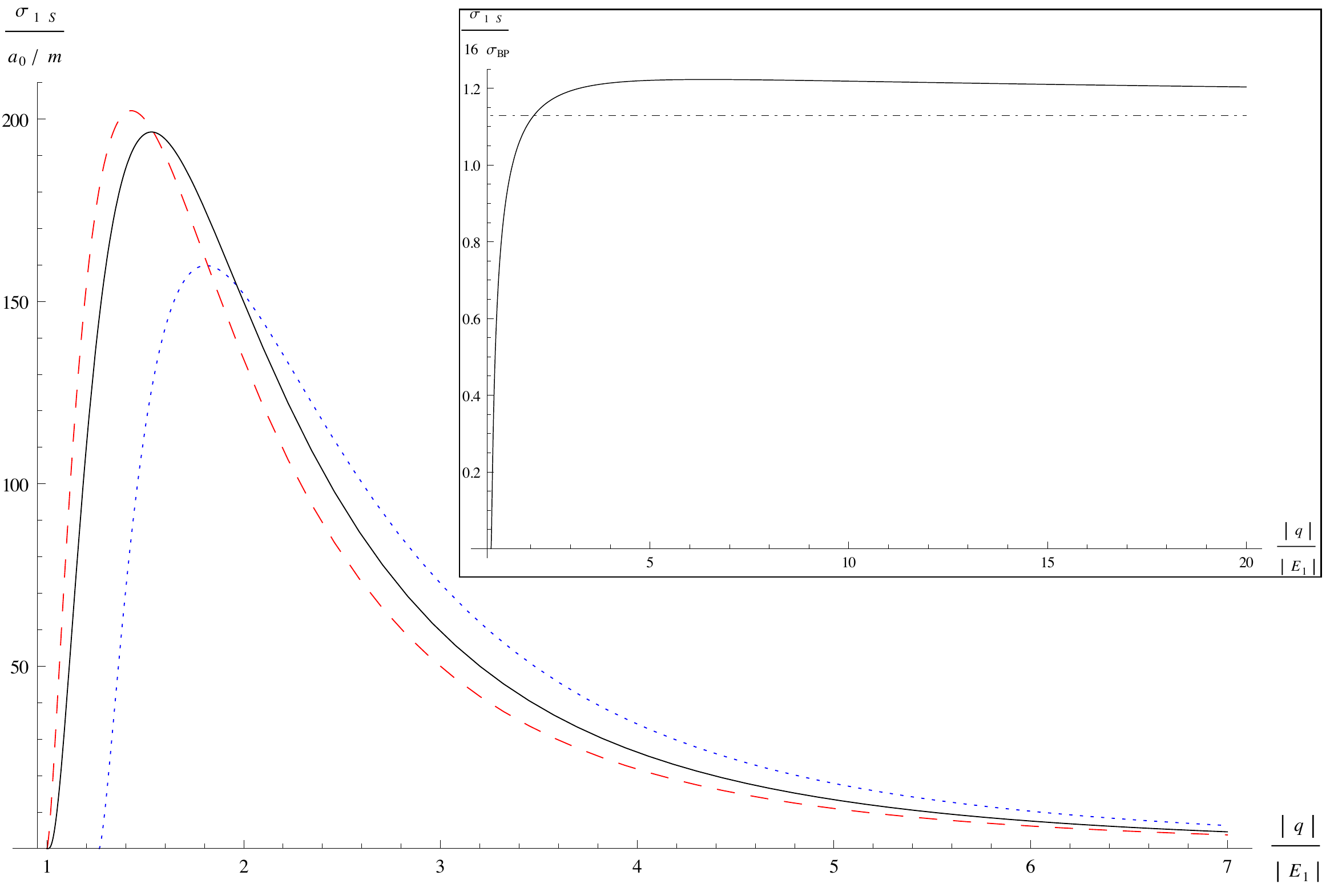}
\end{center}
\caption{In the main plot, the full cross section
$\sigma_{1s}$, given in Eq.~\eqref{crossEFT}, is plotted in continuous
black. The BP cross section, $16\,\sigma_{1S,\mathrm{BP}}$,
Eq. \eqref{eftcrosslargenc}, with the substitution
$\epsilon_{1,\mathrm{BP}}\to\vert E_1\vert$, is plotted in dashed red.
The inset plot shows the ratio $\sigma_{1S}/(16\,\sigma_{1S,\mathrm{BP}})$.
The horizontal dot-dashed line is the asymptotic limit $(17/16)^2$, which
is reached from above. The dotted blue curve in the main plot is the BP
cross section without the substitution $\epsilon_{1,\mathrm{BP}}\to\vert E_1\vert$.}
\label{fig:cross}
\end{figure}

In order to estimate the approximation introduced by ignoring the
octet potential, in Fig.~\ref{fig:cross}, we plot with a continuous black line  
the exact cross section \eqref{crossEFT} and with a dashed red line the BP cross
section \eqref{eftcrosslargenc} as functions of the gluon momentum $\mbq$. 
For a meaningful comparison, for the latter we have performed the substitution 
$\epsilon_{1,\mathrm{BP}}\to\vert E_1\vert$, which guarantees  
that the binding energies are the same in the two cases.
The dotted blue line is the BP cross section \eqref{eftcrosslargenc} 
without this substitution, i.e. in terms  of $\epsilon_{1,\mathrm{BP}}$, 
which shows a larger threshold and a smaller peak. 
In the inset plot, we show the ratio, $\sigma_{1S}/(16\,\sigma_{1S,\mathrm{BP}})$,
of the continuous black and dashed red curves:
the horizontal dot-dashed line is the asymptotic value of $(2+\rho)^2/4$, which, for $\nc=3$, 
yields $(17/16)^2$.

\begin{figure}[ht]
\begin{center}
\includegraphics{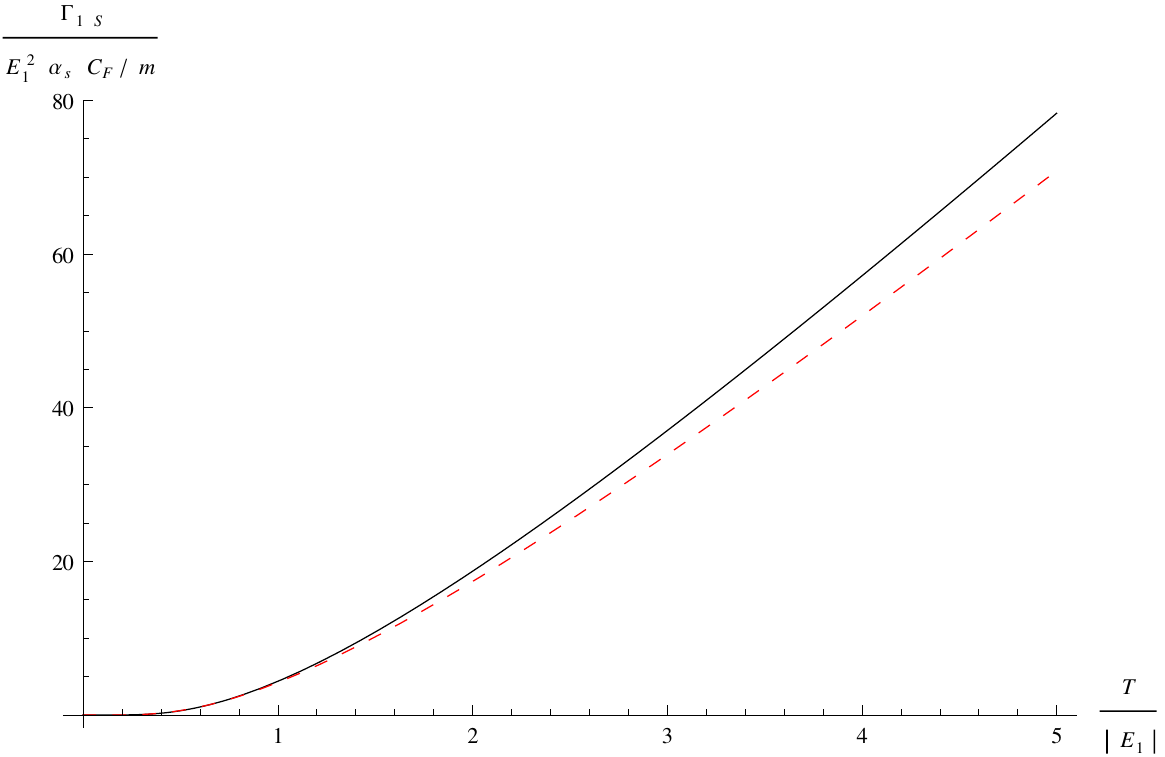}
\end{center}
\caption{The width $\Gamma_{1S}$ is shown as a continuous black line and the width $\Gamma_{1S,\mathrm{BP}}$ as a dashed red line.}
\label{fig_width}
\end{figure}

In Fig.~\ref{fig_width}, we plot the widths $\Gamma_{1S}$ and
$\Gamma_{1S,\mathrm{BP}}$ as a function of the temperature. They have been 
obtained by numerical integration of Eqs.  \eqref{widthinsert} and
\eqref{peskinEFT} respectively. In the latter case, we perform the same
substitution as above, i.e. $\epsilon_{1,\mathrm{BP}}\to\vert E_1\vert$. 
$\Gamma_{1S}$ is the continuous black line, whereas $\Gamma_{1S,\mathrm{BP}}$ is the dashed red line. 
One clearly sees how the full result $\Gamma_{1S}$ overpowers the old BP result 
in whole range $T\simg \vert E_1\vert$, and how the two widths quickly reach an asymptotic linear
regime for $T\gg\vert E_1\vert$, as predicted by the analytical result \eqref{analyticwidth}.
For $T\siml |E_1|$ the two widths become exponentially small.

\begin{figure}[ht]
\begin{center}
\includegraphics[width=12cm]{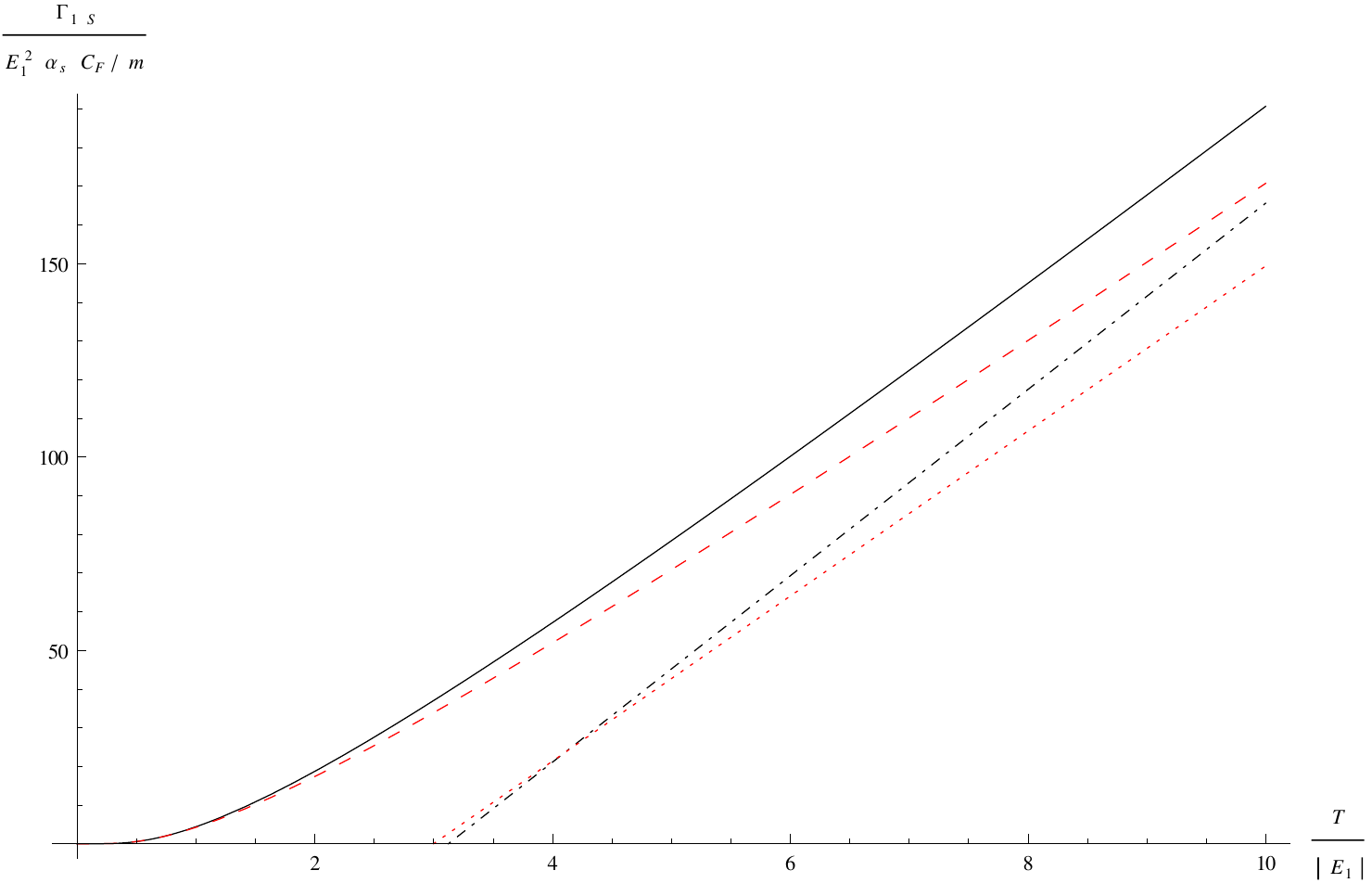}
\end{center}
\caption{The width $\Gamma_{1S}$ is shown as a continuous black line 
  and the corresponding analytical result for $T\gg\vert E_1\vert$, 
  Eq.~\eqref{widthanalytic} with $\rho=1/8$, is plotted as a dot-dashed
  black line. Similarly, the width $\Gamma_{1S,\mathrm{BP}}$ 
  is plotted as a dashed red line and the corresponding analytical result for $T\gg\vert E_1\vert$,
  Eq.~\eqref{widthanalytic} with $\rho=0$, as a dotted red line.}
\label{fig_peskin}
\end{figure}

\begin{figure}[ht]
\begin{center}
\includegraphics[width=12cm]{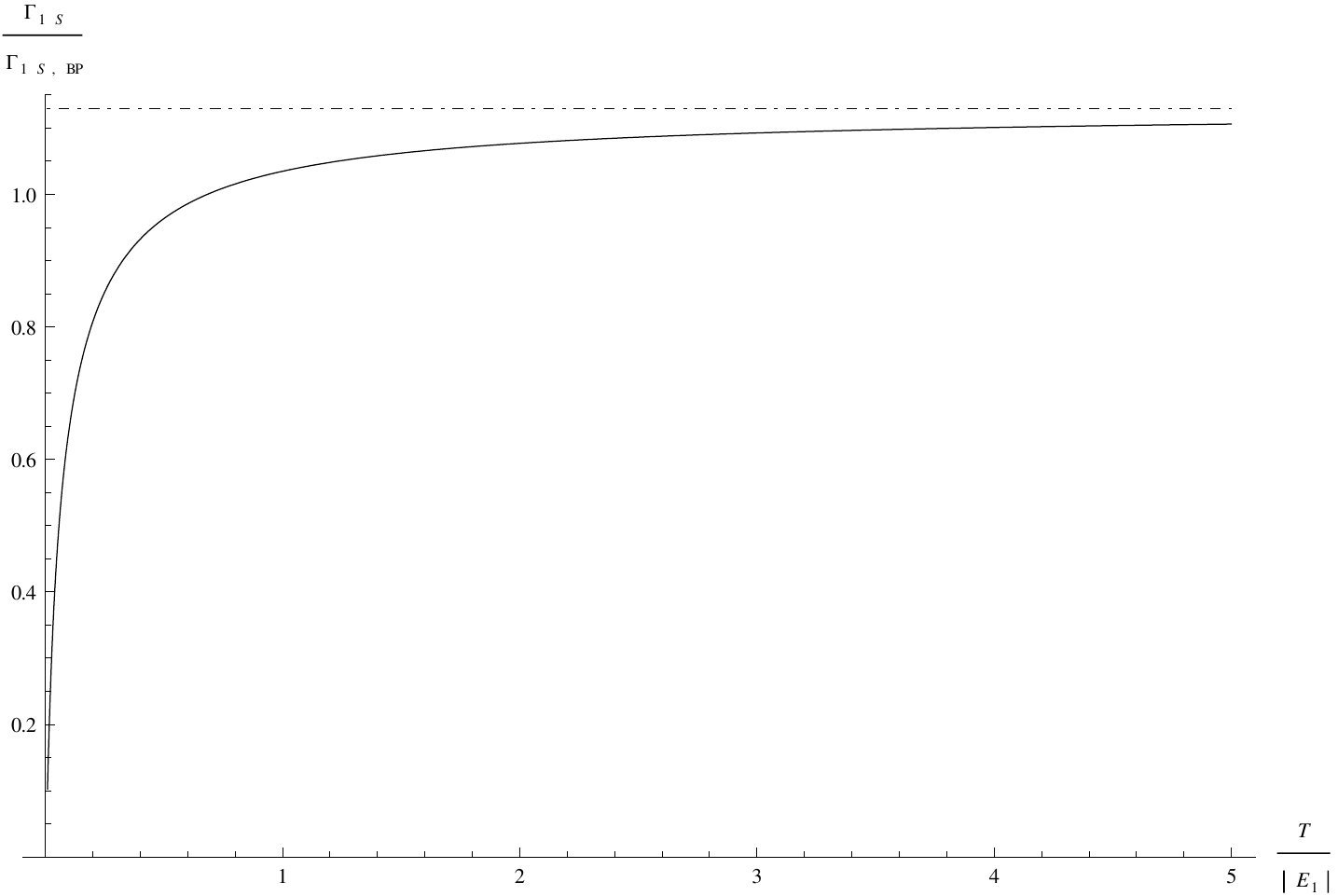}
\end{center}
\caption{Plot of the ratio $\Gamma_{1S}/\Gamma_{1S,\mathrm{BP}}$; the horizontal dot-dashed 
line is the asymptotic limit $(17/16)^2$.}
\label{fig_ratio}
\end{figure}

The analytical, asymptotic expression of the full width  $\Gamma_{1S}$ for $T\gg\vert E_1\vert$
is in Eq.~\eqref{analyticwidth}. In terms of $\rho$, it reads  
\begin{equation}
\label{widthanalytic}
\frac{\Gamma_{1S}}{E_1^2\cf\als/m}=
\frac{16}{3}\left[(2+\rho)^2\frac{T}{\vert E_1\vert}-(2+\rho)^2(3+\rho)\right]+\order{\frac{\vert E_1\vert}{T}}.
\end{equation}
In Fig.~\ref{fig_peskin}, the width \eqref{widthanalytic} for $\nc=3$ ($\rho=1/8$), 
corresponding to the inclusion of the octet potential, is plotted as a dot-dashed black line.
The corresponding plot of $\Gamma_{1S}$, obtained from a numerical integration of
Eq.~\eqref{widthinsert}, is the continuous black line.
The width \eqref{widthanalytic} for $\nc\to\infty$ ($\rho=0$), corresponding to the BP 
approximation of a vanishing octet potential, is plotted as a dotted red line.
The corresponding plot of  $\Gamma_{1S,\mathrm{BP}}$, obtained from a numerical integration of
Eq.~\eqref{peskinEFT}, is the dashed red line.
Both $\Gamma_{1S}$ and $\Gamma_{1S,\mathrm{BP}}$ approach their asymptotic linear regimes 
starting from $T\approx 4\vert E_1\vert$. For  $T\approx 4\vert E_1\vert$, $\Gamma_{1S}$ is 
still larger than its large $T$ asymptotic value by about a factor 2.7.
In Fig.~\ref{fig_ratio}, we plot the ratio $\Gamma_{1S}/\Gamma_{1S,\mathrm{BP}}$, which also shows the 
deviation from the asymptotic limit of $(17/16)^2\approx 1.13$.

Finally, we remark that our results are valid as long as 
$m\als \gg T$ (see Eq. \eqref{scales}), which guarantees that 
the interaction of the bound state with the thermal gluons can be treated as a
chromoelectric dipole. Therefore the regions of the plots where  $T/\vert E_1\vert \simg 1/\als \approx 5$
are to be intended for illustration purposes only.

\section{Conclusions}
\label{sec_concl}
In this letter, we have shown that, under the scale hierarchy \eqref{scales}, 
the leading contribution to the thermal width of a quarkonium $1S$ state 
may be written as a convolution integral of the gluon distribution function 
and a cross section $\sigma_{1S}$, see Eq. \eqref{gluodiss}.
The underlying process is known as singlet-to-octet thermal break-up in the EFT 
literature and as quarkonium thermal gluo-dissociation in the phenomenological 
literature: a colour-singlet $Q\overline{Q}$ state interacts with a gluon of the 
thermal bath and breaks up in an unbound colour-octet  $Q\overline{Q}$ pair.
The cross section $\sigma_{1S}$ can be identified with the in vacuum gluo-dissociation cross section. 
We have derived its explicit expression in Eq. \eqref{crossEFT}. 
This expression includes, for the first time, the contribution of the colour-octet 
potential, i.e. the final-state interactions between the heavy quark and antiquark.
We have shown that the gluo-dissociation cross section and decay width 
reduce to the well-known Bhanot--Peskin result if colour-octet effects 
are ignored, see Eqs. \eqref{eftcrosslargenc} and \eqref{peskinEFT}.
Under the condition $T\gg m\als^2$, the thermal decay width may be expanded 
to give back the singlet-to-octet break-up width calculated in \cite{Brambilla:2010vq}, 
see Eq. \eqref{analyticwidth}.

\begin{figure}[ht]
\begin{center}
\includegraphics[width=12cm]{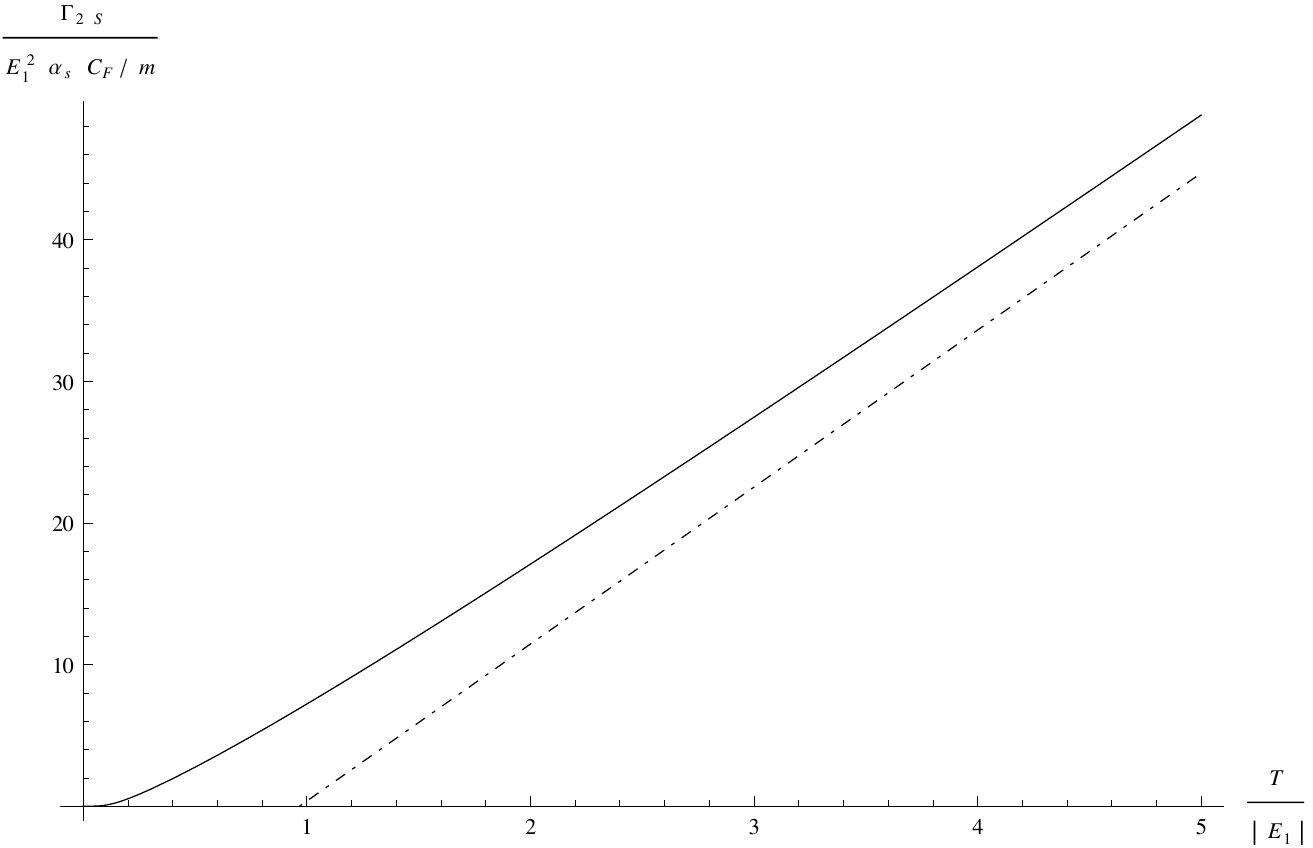}
\end{center}
\caption{The width $\Gamma_{2S}$ is shown as a continuous line as a function of $T/\vert E_1\vert$. The corresponding analytical result for $T\gg\vert E_2\vert$, as obtained in \cite{Brambilla:2010vq}, is plotted as a dot-dashed line.}
\label{fig2s}
\end{figure}

The EFT framework, in which the factorization formula and the cross-section expression 
have been derived, makes clear the region of validity and the accuracy of the 
obtained results. They hold under the conditions  $m\als \gg T$, 
so that the interaction of the bound state with the gluons of the
medium can be described by a chromoelectric dipole interaction, 
$m\als^2 \gg m_D$ and $T \gg m_D$, so that, in first approximation,  
the thermal masses of the gluons can be neglected.
The factorization holds at leading order in the EFT power counting.
Beyond leading order, which includes contributions coming from hard thermal loop resummed gluon propagators, 
the simple convolution formula \eqref{gluodiss} will break down.
The explicit expression of the gluo-dissociation cross section that we 
have presented, as well as the old Bhanot--Peskin expression, 
follows from the assumption that the  $Q\overline{Q}$ pair 
is weakly coupled so that the colour-singlet bound state and the 
colour-octet unbound quark-antiquark pair may be described 
in terms of Coulombic bound or scattering states respectively.
This assumption is likely to hold only for the quarkonium ground state; 
for this reason, we have restricted our analysis to the dissociation 
of quarkonium $1S$ states, although an extension to quarkonium states 
with arbitrary quantum numbers would be straightforward 
(the cross section for an arbitrary state can be
found making the substitutions $E_1\to E_n$,
$\langle 1S|\to \langle nll_z|$ and
$|{\bf p} 1\rangle \to \sum_l|{\bf p} l\rangle$
in Eq.~\eqref{sigma}, see footnote \ref{foot} for
the explicit expression of the $\sigma_{2S}$ cross section and 
Fig.~\ref{fig2s} for the $\Gamma_{2S}$ width). 
In view of this, we stress that phenomenological 
studies that make use of the Bhanot--Peskin gluo-dissociation formula,  
but fix the binding energy by some non-Coulombic model of the bound state, 
cannot be justified within QCD.

The gluo-dissociation cross section is shown in Fig.~\ref{fig:cross}. 
The impact of the colour-octet effects is dramatic for gluon momenta, 
$\bq$, close to the threshold: the full cross section \eqref{crossEFT} falls 
off exponentially while the BP cross section \eqref{eftcrosslargenc}
falls off like $(\mbq-\epsilon_{1,\mathrm{BP}})^{3/2}$. 
At larger gluon momenta, if the thresholds in both cross sections are chosen to be equal to $|E_1|$,
then for $\mbq = 2 |E_1|$ the full cross section is larger than the BP one 
by about 12\% and for $\mbq = 5 |E_1|$ by about 22\%.
In the asymptotic high-momentum limit, the full cross section overshoots the BP 
one by about 13\%. The thermal decay width is shown as a function of the 
temperature in Fig.~\ref{fig_width}. Much larger temperatures than those plotted  
would likely violate the bound  $m\als \gg T$. At temperatures lower than the 
energy $|E_1|$ the full thermal width falls off faster than the BP one. 
For $T=0.3 |E_1|$, the full width is smaller than the BP width 
by about 12\%, while for $T=|E_1|$, the full width is larger than the BP width 
by about 3\% and for $T=5 |E_1|$ it is larger by about 11\%.
According to \cite{Vairo:2010bm}, the region $|E_1| \ge T \ge 0.3|E_1|$ may be 
of relevance for $\Upsilon(1S)$ produced in heavy-ion experiments at LHC.

\section*{Note added}
While this paper was in the final writing up, a paper
appeared \cite{Brezinski:2011ju} where the gluo-dissociation cross
section in the presence of the octet potential was also obtained. The
result is shown in Eq.~(4) and for $n=1$, after correcting some typos, 
agrees with our Eq.~\eqref{crossEFT}.

\section*{Acknowledgements}
N.B. and A.V. thank Hossein Malekzadeh for collaboration 
during the early stages of this work. N.B., J.G. and A.V. thank Ralf
Rapp for discussions. M.A.E acknowledges useful discussions with Xingbo
Zhao. We acknowledge financial support from the DFG project BR4058/1-1
``Effective field theories for strong interactions with heavy
quarks''. N.B., J.G. and A.V.  acknowledge financial support from the
DFG cluster of excellence ``Origin and structure of the universe''
(\href{http://www.universe-cluster.de}{www.universe-cluster.de}).


\begin{thebibliography}{99}
\bibitem{Matsui:1986dk}
  T.~Matsui and H.~Satz,
  Phys.\ Lett.\  B {\bf 178} (1986) 416.

\bibitem{Brambilla:2004wf}
  N.~Brambilla {\it et al.},
  Heavy quarkonium physics,
  CERN-2005-005, (CERN, Geneva, 2005)
  [arXiv:hep-ph/0412158].

\bibitem{Brambilla:2010cs}
  N.~Brambilla {\it et al.},
  Eur.\ Phys.\ J.\  C {\bf 71} (2011) 1534
  [arXiv:1010.5827 [hep-ph]].

\bibitem{Laine:2006ns}
  M.~Laine, O.~Philipsen, P.~Romatschke and M.~Tassler,
  JHEP {\bf 0703} (2007) 054 
  [arXiv:hep-ph/0611300].

\bibitem{Laine:2007gj}
  M.~Laine,
  JHEP {\bf 0705} (2007) 028 
  [arXiv:0704.1720 [hep-ph]].

\bibitem{Burnier:2007qm}
  Y.~Burnier, M.~Laine and M.~Vepsalainen,
  JHEP {\bf 0801} (2008) 043 
  [arXiv:0711.1743 [hep-ph]].

\bibitem{Beraudo:2007ky}
  A.~Beraudo, J.~P.~Blaizot and C.~Ratti,
  Nucl.\ Phys.\  A {\bf 806} (2008) 312 
  [arXiv:0712.4394 [nucl-th]].

\bibitem{Escobedo:2008sy}
  M.~A.~Escobedo and J.~Soto,
 Phys. Rev. A {\bf 78} (2008) 032520 
 [arXiv:0804.0691 [hep-ph]].

\bibitem{Escobedo:2010tu}
  M.~A.~Escobedo and J.~Soto,
  Phys.\ Rev.\  A {\bf 82} (2010) 042506 
  [arXiv:1008.0254 [hep-ph]].

\bibitem{Escobedo:2011ie}
  M.~A.~Escobedo, J.~Soto and M.~Mannarelli,
  Phys.\ Rev.\  D {\bf 84} (2011) 016008
  [arXiv:1105.1249 [hep-ph]].

\bibitem{Brambilla:2008cx}
  N.~Brambilla, J.~Ghiglieri, A.~Vairo and P.~Petreczky,
  Phys.\ Rev.\  D {\bf 78} (2008) 014017
  [arXiv:0804.0993 [hep-ph]].

\bibitem{Brambilla:2010vq}
  N.~Brambilla, M.~A.~Escobedo, J.~Ghiglieri, J.~Soto and A.~Vairo,
  JHEP {\bf 1009} (2010) 038
  [arXiv:1007.4156 [hep-ph]].

\bibitem{Brambilla:2011mk}
  N.~Brambilla, M.~A.~Escobedo, J.~Ghiglieri and A.~Vairo,
  JHEP {\bf 1107} (2011) 096
  [arXiv:1105.4807 [hep-ph]].

\bibitem{Mocsy:2007jz}
  A.~Mocsy, P.~Petreczky,
  Phys.\ Rev.\ Lett.\  {\bf 99 } (2007)  211602.
  [arXiv:0706.2183 [hep-ph]].

\bibitem{Rapp:2008tf}
  R.~Rapp, D.~Blaschke and P.~Crochet,
  Prog.\ Part.\ Nucl.\ Phys.\  {\bf 65} (2010) 209
  [arXiv:0807.2470 [hep-ph]].

\bibitem{Kluberg:2009wc}
  L.~Kluberg and H.~Satz,
  arXiv:0901.3831 [hep-ph].

\bibitem{Kharzeev:1994pz}
  D.~Kharzeev and H.~Satz,
  Phys.\ Lett.\  B {\bf 334} (1994) 155
  [arXiv:hep-ph/9405414].

\bibitem{Xu:1995eb}
  X.~M.~Xu, D.~Kharzeev, H.~Satz and X.~N.~Wang,
  Phys.\ Rev.\  C {\bf 53} (1996) 3051
  [arXiv:hep-ph/9511331].
  
\bibitem{Xu:2007yu}
  X.~-M.~Xu,
  Nucl.\ Phys.\  {\bf A658 } (1999)  165
  [arXiv:0704.0668 [hep-ph]].

\bibitem{Polleri:2003kn}
  A.~Polleri, T.~Renk, R.~Schneider, W.~Weise,
  Phys.\ Rev.\  {\bf C70 } (2004)  044906
  [arXiv:nucl-th/0306025 [nucl-th]].

\bibitem{Patra:2004wf}
  B.~K.~Patra, V.~J.~Menon,
  Eur.\ Phys.\ J.\  {\bf C37 } (2004)  115 
  [nucl-th/0401025].

\bibitem{Patra:2005yg}
  B.~K.~Patra, V.~J.~Menon,
  Eur.\ Phys.\ J.\  {\bf C44 } (2005)  567
  [nucl-th/0503034].

\bibitem{Patra:2005bi}
  B.~K.~Patra, V.~J.~Menon,
  Eur.\ Phys.\ J.\  {\bf C48 } (2006)  207
  [nucl-th/0512103].

\bibitem{Wong:2004zr}
  C.~-Y.~Wong,
  Phys.\ Rev.\  {\bf C72 } (2005)  034906
  [hep-ph/0408020].

\bibitem{Arleo:2004ge}
  F.~Arleo, J.~Cugnon, Y.~Kalinovsky,
  Phys.\ Lett.\  {\bf B614 } (2005)  44
  [hep-ph/0410295].

\bibitem{Thews:2005vj}
  R.~L.~Thews, M.~L.~Mangano,
  Phys.\ Rev.\  {\bf C73 } (2006)  014904
  [nucl-th/0505055].

\bibitem{Grandchamp:2005yw}
  L.~Grandchamp, S.~Lumpkins, D.~Sun, H.~van Hees, R.~Rapp,
  Phys.\ Rev.\  {\bf C73 } (2006)  064906
  [hep-ph/0507314].

\bibitem{Park:2007zza}
  Y.~Park, K.~I.~Kim, T.~Song, S.~H.~Lee and C.~Y.~Wong,
  Phys.\ Rev.\  C {\bf 76} (2007) 044907
  [arXiv:0704.3770 [hep-ph]].

\bibitem{Liu:2009wza}
  Y.~Liu, Z.~Qu, N.~Xu, P.~Zhuang,
  J.\ Phys.\ G {\bf G37 } (2010)  075110.
  [arXiv:0907.2723 [nucl-th]].

\bibitem{Qu:2009sk}
  Z.~Qu, Y.~Liu, N.~Xu, P.~Zhuang,
  Nucl.\ Phys.\  {\bf A830 } (2009)  335C
  [arXiv:0907.3626 [nucl-th]].

\bibitem{Zhou:2009vz}
  K.~Zhou, N.~Xu, P.~Zhuang,
  Nucl.\ Phys.\  {\bf A834 } (2010)  249C
  [arXiv:0911.5008 [nucl-th]].

\bibitem{Song:2010ix}
  T.~Song, W.~Park, S.~H.~Lee,
  Phys.\ Rev.\  {\bf C81 } (2010)  034914
  [arXiv:1002.1884 [nucl-th]].

\bibitem{Zhao:2010nk}
  X.~Zhao and R.~Rapp,
  Phys.\ Rev.\  C {\bf 82} (2010) 064905
  [arXiv:1008.5328 [hep-ph]].

\bibitem{Liu:2010ej}
  Y.~Liu, B.~Chen, N.~Xu, P.~Zhuang,
  Phys.\ Lett.\  {\bf B697 } (2011)  32
  [arXiv:1009.2585 [nucl-th]].

\bibitem{Uphoff:2011fu}
  J.~Uphoff, K.~Zhou, O.~Fochler, Z.~Xu, C.~Greiner,
  [arXiv:1104.2437 [hep-ph]].

\bibitem{Mandal:2011jx}
  M.~Mandal, P.~Roy,
  [arXiv:1105.5528 [hep-ph]]. 
  
  \bibitem{Rapp:2009my}
  R.~Rapp, H.~van Hees,
  [arXiv:0903.1096 [hep-ph]].

\bibitem{Grandchamp:2001pf}
  L.~Grandchamp, R.~Rapp,
  Phys.\ Lett.\  {\bf B523 } (2001)  60
  [hep-ph/0103124].

 \bibitem{preparation}
N.~Brambilla, M.~A.~Escobedo, J.~Ghiglieri and A.~Vairo, TUM-EFT 27/11, in preparation.

\bibitem{Peskin:1979va}
  M.~E.~Peskin,
  Nucl.\ Phys.\  B {\bf 156} (1979) 365.

\bibitem{Bhanot:1979vb}
  G.~Bhanot and M.~E.~Peskin,
  Nucl.\ Phys.\  B {\bf 156} (1979) 391.

\bibitem{PHDJacopo}
J.~Ghiglieri, Ph.D. Thesis, TU Munich (July, 2011).

\bibitem{Vairo:2010bm}
  A.~Vairo,
  AIP Conf.\ Proc.\  {\bf 1317 } (2011)  241-249.
  [arXiv:1009.6137 [hep-ph]].

\bibitem{Chatrchyan:2011pe}
  S.~Chatrchyan {\it et al.}  [CMS Collaboration],
  Phys.\ Rev.\ Lett.\  {\bf 107} (2011) 052302
  [arXiv:1105.4894 [nucl-ex]],
  and CMS-PAS-HIN-10-006.

\bibitem{Strickland:2011mw}
  M.~Strickland,
  arXiv:1106.2571 [hep-ph].

\bibitem{Caswell:1985ui}
W.~E.~Caswell and G.~P.~Lepage,
Phys.\ Lett.\ B {\bf 167} (1986) 437. 

\bibitem{Bodwin:1994jh}
G.~T.~Bodwin, E.~Braaten and G.~P.~Lepage,
Phys.\ Rev.\ D {\bf 51} (1995) 1125 
[Erratum-ibid.\ D {\bf 55} (1997) 5853].

\bibitem{Pineda:1997bj}
  A.~Pineda and J.~Soto,
  Nucl.\ Phys.\ Proc.\ Suppl.\  {\bf 64} (1998) 428 
  [arXiv:hep-ph/9707481].

\bibitem{Brambilla:1999xf}
  N.~Brambilla, A.~Pineda, J.~Soto and A.~Vairo,
  Nucl.\ Phys.\  B {\bf 566} (2000) 275 
  [arXiv:hep-ph/9907240].

\bibitem{Vairo:2009ih}
  A.~Vairo,
  PoS {\bf CONFINEMENT8 } (2008)  002.
  [arXiv:0901.3495 [hep-ph]].

\bibitem{Aarts:2011sm}
  G.~Aarts, C.~Allton, S.~Kim, M.~P.~Lombardo, M.~B.~Oktay, S.~M.~Ryan, D.~K.~Sinclair, J.~-I.~Skullerud,
  [arXiv:1109.4496 [hep-lat]].
  
\bibitem{Kniehl:1999ud}
  B.~A.~Kniehl and A.~A.~Penin,
  Nucl.\ Phys.\  B {\bf 563} (1999) 200
  [arXiv:hep-ph/9907489].

\bibitem{Kniehl:2002br}
  B.~A.~Kniehl, A.~A.~Penin, V.~A.~Smirnov and M.~Steinhauser,
  Nucl.\ Phys.\  B {\bf 635} (2002) 357
  [arXiv:hep-ph/0203166].

\bibitem{Kobes:1986za}
  R.~L.~Kobes, G.~W.~Semenoff,
  Nucl.\ Phys.\  {\bf B272 } (1986)  329.

\bibitem{abramovitz+stegun}
M.~Abramovitz and I.~A.~Stegun,
Handbook of Mathematical Functions with Formulas, Graphs, and Mathematical Tables, Dover (1964).

\bibitem{Brezinski:2011ju}
  F.~Brezinski, G.~Wolschin,
  [arXiv:1109.0211 [hep-ph]].

\end{thebibliography}
\end{document}